\documentclass[twocolumn,amsmath,floatfix, aps,amssymb,superscriptaddress,showpacs]{revtex4}
\usepackage{graphicx}
\usepackage{dcolumn}
\usepackage{setspace}
\begin{document}
\title{Memory-induced anomalous dynamics: emergence of diffusion, subdiffusion, and superdiffusion 
from a single random walk model}
\author{ Niraj Kumar}
\affiliation{Department of Chemistry and Biochemistry, and BioCircuits Institute, University of
California San Diego, La Jolla, California 92093-0340, USA}
\author{Upendra Harbola}
\affiliation{Department of Chemistry and Biochemistry, and BioCircuits Institute, University of
California San Diego, La Jolla, California 92093-0340, USA}
\author{Katja Lindenberg}
\affiliation{Department of Chemistry and Biochemistry, and BioCircuits Institute, University of
California San Diego, La Jolla, California 92093-0340, USA}

\begin{abstract}
We present a random walk model that exhibits asymptotic subdiffusive, diffusive, and superdiffusive
behavior in different parameter regimes. This appears to be the first instance of a single
random walk model leading to all three forms of behavior by simply changing parameter values. 
Furthermore, the model offers the great advantage of analytic tractability.
Our model is non-Markovian in that the next jump of the walker is (probabilistically) determined
by the history of past jumps. It also has elements of intermittency in that one possibility at each
step is that the walker does not move at all.  This rich encompassing scenario arising from a single model
provides useful insights into the source of different types of asymptotic behavior. 

\end{abstract}


\maketitle{}

\section{Introduction}
\label{introduction}

The use of random walk models in statistical physics dates back to the very earliest days of the
subject~\cite{Ba1900,Ei1905,Ch43,Hughes95,Ma99,AvHa00}.
They have been used so broadly and pervasively that it is impossible to make proper
reference to so big a subject.  The original random walk models were typically ones in which the
walker moves via a series of random transitions characterized by finite length scales for each step
and finite time scales between transitions.  As long as these general features characterize the
walk, the mean square displacement of a walker from its point of origin in the absence of a bias,
calculated as an average over many randomly generated trajectories, grows linearly with time.  If
there is a bias, then the mean square displacement around the average trajectory grows linearly with
time.  This linear growth has become the universal identifier of what is known as ``normal
transport."  Diffusion is the quintessential macroscopic normal transport mechanism, and is often
arrived at by taking appropriate long-time and short-distance limits of random
walks~\cite{Hughes95,Mainardi}.

It is then not surprising that any random walk (or, for that matter, any stochastic process) that
leads to a mean square displacement that does not grow linearly with time is called ``anomalous."
This characterization is shared by stochastic processes whose mean square displacement grows either
sublinearly or superlinearly with time, the former often called ``subdiffusive" and the latter
``superdiffusive" processes.   In the world of random walks, subdiffusive behavior is often associated
with waiting time distributions between steps that have fat tails, e.g., that decay as an inverse
small power of time~\cite{MetzlerKlafterPhysReport,MetzlerKlafterRestaurant,Klages}. 
The average time between transitions in this case is infinite.  On the other
hand, superdiffusive behavior is typically associated with step length distributions that have fat
tails, so that the average distance per step is infinite~\cite{Klages}.  L\'evy flights are well-known examples of
such processes~\cite{Klages}.  

In recent years there has been considerable interest in formulating stochastic models that can
exhibit different types of behavior in different parameter regimes.  More specifically, for example
in the context of random walks, it has been observed in a variety of contexts that the behavior of
a walker may be normal in some parameter regimes but anomalous in others.  This is almost trivial to
envision in terms of the quantities already introduced above.  For instance,
consider a model with a waiting time distribution with a power law tail of the form $t^{-\alpha}$.
It may then happen that in some regimes of the model the waiting time distribution decays slowly
($\alpha<2$), so that the average waiting time between transitions diverges.  In a different regime
of the model the decay of the waiting time may be sufficiently rapid ($\alpha>2$) for there to be a
finite mean transition time.  The process would be subdiffusive in the first regime and normal in the
second. On the other hand, for example if a model has a jumping pattern with an infinite average jump
distance in some regime but a finite one in another, the model would exhibit superdiffusive behavior
in the first and normal diffusive behavior in the second.  Transitions between normal and anomalous
behavior within a model can occur in many other ways as well. The literature is far too large
to reference it properly. Perhaps the earliest of these in the random walk context was introduced by
Zumofen and Klafter~\cite{Zumofen} to describe the dynamics generated by iterated maps.
A more recent example directly
related to the discussion in our paper deals with a class of random walk models called ``elephant
random walks," a term coined by its creators~\cite{ScTr04} because they involve a sort of perfect
memory often (and most likely inappropriately) associated with elephants. In these models, which have
also been mentioned in applications as diverse as ecology, economic data, and DNA strings,
the memory induces a transition between normal and
superdiffusive behavior observed with a change in parameter
values~\cite{CrSiVi07,Kenkre07,FeCrViAl10,HaTo09,BoDaFr08,PaEs06,IsKoIn05}.

While there is thus a plenitude of stochastic models in which the mean square displacement changes
from one behavior to another as a parameter of the model is varied, it is more difficult to find
models that exhibit all three forms of behavior, that is, ones in which a change of parameters
leads to asymptotic subdiffusive, normal, and superdiffusive behavior. Such versatility can be found
in generalized Langevin equations~\cite{Porra} and in dynamics governed by fractional Brownian
motion or fractional Langevin equations (a helpful list of references to the fractional Langevin
literature can be found in~\cite{Metzler}). 
However, to the best of our knowledge, there is no single \emph{random walk} model
that exhibits all three forms of behavior.  In this
paper we present such a model. It is a random walk model with a memory, and a simple sweep in parameter
values indicating the direction of motion of the next change can lead to transitions covering all
three forms of behavior. The model is inspired by the ``elephant random walk"
introduced in~\cite{ScTr04}.
In Sec.~\ref{model} we present the random walk model. In Sec.~\ref{moments} we calculate the moments of interest
to characterize the nature of the motion of the random walker. A detailed analysis of the results is
presented in Sec.~\ref{analysis}. We conclude with a brief recap in Sec.~\ref{conclusion}.

\section{The model}
\label{model}

Following the notation in~\cite{ScTr04}, consider a random walker on a one-dimensional infinite
lattice with unit distance between adjacent
lattice sites.  Steps occur at discrete time intervals. At each step the walker can take one of
three actions: it can move to the nearest neighbor site to its right, to the site on its left, or
it can remain at its present location (that is, the walk is intermittent).  We denote the
position of the walker at time step $t$ as $x_t$. The position of the walker
at time step $t+1$ is
\begin{equation}{\label{n_0}}
x_{t+1}=x_t+\sigma_{t+1}.
\end{equation}
Here $\sigma_{t+1}$ is a random number which 
can take on one of the values $-1$, $0$ or $+1$. The choice of this random number at each step depends on the
entire history of the walk, $\{\sigma_t\}=(\sigma_1,...,\sigma_t)$, as follows. A random previous time $k$ 
between $1$ and $t$ is chosen with uniform probability.
If $\sigma_k=\pm 1$, with probability $p$ the walker takes the same step at time $t+1$,
i.e., $\sigma_{t+1}=\sigma_k$. With probability $q$, the walker takes the opposite action,
$\sigma_{t+1}=-\sigma_k$.  The walker can stay at rest with probability $r$, and of course
$p+q+r=1$.  If $\sigma_k=0$, the walker stays at rest 
with probability $1$.  The process is started at time $t=1$ by allowing the walker to move 
to the right with probability $s$ and to the left with probability $1-s$, i.e., the first
step excludes the possibility
that the walker may not move. If the walker is initially at $x=0$,  then the position of the 
walker at time $t>1$ is given by
\begin{equation}{\label{e2s_1}}
x_t=\sum_{k=1}^{t}\sigma_k.
\end{equation}
We make a special point here about the significance of the parameters $p$ and $q$: they are
{\bf not} the familiar parameters that indicate a next-step asymmetry. Rather, they are memory
parameters that indicate whether the walker, if it moves at all at the next step, is likely to follow
the randomly chosen past step (persistence probability $p$) or is, instead, a rebellious walker
who does the opposite (probability $q$). It is thus not straightforward to a priori predict the
direction and growth properties
of the walk as a function of these parameters.  For $r=0$, the model reduces to that of~\cite{ScTr04}.
However, we will show that the inclusion of a probability that the walker neither follows nor rebels
against the randomly chosen prior step, $r\neq 0$, turns out to be the crucial generalization that
leads to the all-encompassing model.

\section{Moments}
\label{moments}

The quantities of interest to characterize the nature of the motion of the walker are the mean
displacement and the mean square displacement as a function of time, both of which we are able
to calculate analytically. The parameters of the problem are two of the stepping probability
parameters; those that turn out to be most useful in characterizing the properties
of the walk are the ``staying probability," $r$, and the memory asymmetry parameter,
\begin{equation}
\gamma= p-q.
\end{equation}

We start by noting that for a given history $\{\sigma_t\}$, the conditional probability
that $\sigma_{t+1}=\sigma$, where $t\ge1$, can be written as,
\begin{eqnarray}
{\label{er_1}}
P[\sigma_{t+1}&=&\sigma|\{\sigma_t\}]= 1-\sigma^2\nonumber\\
&&+ \frac{1}{2t}\sum_{k=1}^{t}\left[\sigma_k^2\left(3\sigma^2-2\right)
\left(1-r\right)+\sigma\sigma_k\gamma\right],
\end{eqnarray}
and for $t=0$,
\begin{equation}{\label{er_2}}
P[\sigma_1=\sigma]=\frac{1}{2}\left(1+(2s-1)\sigma\right).
\end{equation}
Using Eq.~(\ref{er_1}), the conditional mean values of $\sigma_{t+1}$ for $t>1$ in 
a given realization is given as
\begin{equation}
\langle \sigma_{t+1}|\{\sigma_t\}\rangle=\sum_{\sigma=\pm1,0}\sigma P[\sigma_{t+1}=\sigma|\{\sigma_t\}]=\frac{\gamma}{t}x_t,
\end{equation}
which, on averaging over all the histories, gives the following 
mean values,
\begin{equation}{\label{er_3}}
\langle\sigma_{t+1}\rangle=\frac{\gamma}{t}\langle x_t\rangle,
\end{equation}
where $\langle x_t\rangle$ is the mean displacement of the 
walker. Performing the average of Eq.~(\ref{n_0}) and using Eq.~(\ref{er_3}) leads to the 
recursive equation 
\begin{equation}{\label{er_4}}
\langle x_{t+1}\rangle=\left(1+\frac{\gamma}{t}\right)\langle x_t\rangle,
\end{equation}
whose solution is~\cite{ScTr04}
\begin{eqnarray}{\label{er_5}}
\langle x_t\rangle&=&(2s-1)\frac{\Gamma(t+\gamma)}{\Gamma(1+\gamma)\Gamma(t)}\nonumber\\\nonumber\\
         &\sim& t^{\gamma}~~~{\rm{for}}~~~ t\gg1.
\end{eqnarray}
The mean position of the walker vanishes if $s=1/2$.  For $s>1/2$ and $s<1/2$, the 
mean position is positive and negative respectively. Thus, the first step, and the first step alone,
determines whether the walker moves to the right or left macroscopically.
The further evolution of the mean displacement does depend on the parameter $\gamma$.
We note that for a symmetric memory, $\gamma=0$
($p=q$), the mean position is time independent. For $\gamma>0$
($p>q$), the mean position increases with time with an exponent which is smaller than unity for
nonzero values of the rebellion parameter $q$, and so the velocity of the walker decreases with
time. For $\gamma<0$ ($p<q$), the mean position decreases monotonically
with time at long times and thus, on average, the walker returns to the origin and
thus remains localized in space.
This type of motion might be useful to model the dynamics of animal home range behavior~\cite{Borger08}. 

We next compute the second moment of the displacement, $\langle x_t^2 \rangle $. For this, using Eq.~(\ref{n_0}), 
we note that
\begin{equation}
 x_{t+1}^2=x_t^2+\sigma_{t+1}^2+2x_t\sigma_{t+1}.
\end{equation}
For a given history $\{\sigma_t\}$, the conditional average of $x_{t+1}^2$ is then
\begin{equation}{\label{n_c}}
 \langle x_{t+1}^2|\{\sigma_t\}\rangle=x_t^2 +\langle\sigma_{t+1}^2|\{\sigma_t\}\rangle+2 x_t \langle\sigma_{t+1}|\{\sigma_t\}\rangle.
\end{equation}
Here we have made use of the fact that for a given history $\{\sigma_t\}$, $x_t$ is known, which allows us
to write $\langle x_t\sigma_{t+1}|\{\sigma_t\}\rangle=  x_t \langle\sigma_{t+1}|\{\sigma_t\}\rangle$. 
Using Eq.~(\ref{er_1}), the conditional mean values for $\sigma_{t+1}^2$ can be written as
\begin{equation}{\label{n_s2}}
 \langle\sigma_{t+1}^2|\{\sigma_t\}\rangle=\frac{1-r}{t}\sum_{k=1}^{t}\sigma_k^2.
\end{equation}
Using Eqs.~(\ref{er_3}) and (\ref{n_s2}) in Eq.~(\ref{n_c}) and averaging over all histories,
we arrive at the recursive relation
\begin{equation}{\label{er_6}}
\langle x_{t+1}^2\rangle=\left(1+\frac{2\gamma}{t}\right)\langle x_t^2\rangle +\frac{1-r}{t}\left\langle \sum_{k=1}^{t}\sigma_k^2\right\rangle.
\end{equation}
For $r=0$, $\sigma_k^2$ is always $1$ and so the term $\langle\sum_{k=1}^{t}\sigma_k^2\rangle$ is
simply $t$.  For $r=1$ the sum is unity because only the first step contributes.
However, for nonzero values of $r$, $\sigma_k^2$ can be either $1$ or $0$.

\begin{figure}
\includegraphics[width=8cm]{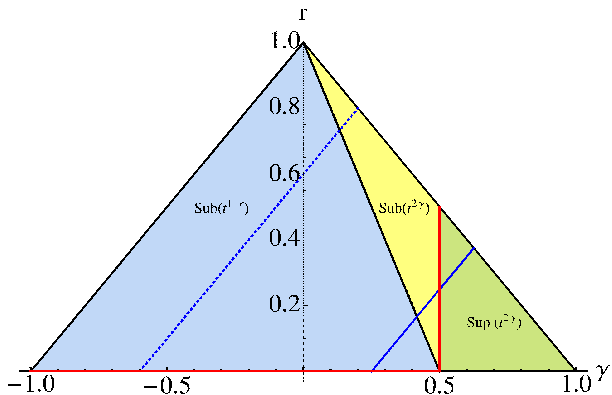}
\caption{(Color online) Phase diagram for the long-time behavior of the variance $Var$ of the random walker.
The leftmost triangle, labeled as Sub$(t^{1-r})$ (light blue in color),
indicates one regime of subdiffusive behavior.  The next
triangular wedge, Sub$(t^{2\gamma})$ (yellow in color) is also a regime of subdiffusive behavior
because here $\gamma< 0.5$ so the time exponent is less than unity. 
The two are separated by the line $r=1-2\gamma$ ($p=3q$).  The last triangular wedge,
Sup$(t^{2\gamma})$ (light green in color) is a superdiffusive regime because here $\gamma>0.5$.
The solid line connecting the subdiffusive and superdiffusive regions (red in color),
as well as the $r=0$ line for $\gamma<1/2$ (also red in color),
indicate the regimes of normal diffusive behavior because here the variance grows as
$Var \sim t^{2\gamma} \sim t$.  The inclined solid and dashed thick lines indicate the variation of the
behavior of the variance for fixed values of the persistence parameter $p$.  The dashed (blue in
color) line is for $p=0.3$ and, as $q$ and $r$ vary for this fixed value of $p$, the walker's behavior
remains subdiffusive throughout, although the form of the exponent changes. The solid (also blue in
color) line is for $p=0.625$.  Now as $q$ and $r$ vary, the walker's behavior covers all three
regimes of behavior: subdiffusive (of two sorts), diffusive, and superdiffusive.}
\label{fig:phase}
\end{figure}

In order to compute $\left\langle \sum_{k=1}^{t}\sigma_k^2\right\rangle$, using Eq.~(\ref{n_s2}) 
 we write
\begin{equation}
\langle\sigma_{t+1}^2|\{\sigma_t\}\rangle=\frac{t-1}{t}\langle\sigma_t^2|\{\sigma_{t-1}\}|\rangle
 +\frac{1-r}{t}\sigma_t^2,
\end{equation}
 which on averaging over all histories gives the recursive relation
\begin{equation}
\langle \sigma_{t+1}^2\rangle=\left(1-\frac{r}{t}\right)\langle\sigma_t^2\rangle,
\end{equation}
leading to the solution
\begin{equation}{\label{n_l}}
 \langle\sigma_t^2\rangle=\frac{\Gamma(t-r)}{\Gamma(1-r)\Gamma(t)}.
\end{equation}
Using Eq.~(\ref{n_l}) in (\ref{n_s2}) gives
\begin{equation}{\label{er_9}}
\left\langle\sum_{k=1}^{t}\sigma_k^2\right\rangle 
=\frac{\Gamma(t-r+1)}{\Gamma(2-r)\Gamma(t)}.
\end{equation}
Substitution into Eq.~(\ref{er_6}) and solution of the recursion equation leads to the mean square
displacement for $\gamma\neq 0$
\begin{eqnarray}
\langle x_t^2 \rangle &=&\frac{1}{(2\gamma+r-1)\Gamma(t)}\left(\frac{\Gamma(t+2\gamma)}
{\Gamma(2\gamma)}-\frac{\Gamma(1+t-r)}{\Gamma(1-r)}\right)\nonumber\\
&\sim&\frac{1}{(2\gamma+r-1)}\left(\frac{t^{2\gamma}}{\Gamma(2\gamma)}-\frac{t^{1-r}}{\Gamma(1-r)}\right),
{\label{er_11}}
\end{eqnarray}
Note that the dominance of the first or second terms is separated by the line $r=1-2\gamma$.
For $\gamma=0$, we find
\begin{equation}
\langle x_t^2\rangle=\frac{\Gamma(1+t-r)}{\Gamma(2-r)\Gamma(t)}
\sim\frac{1}{\Gamma(2-r)}t^{1-r}.
{\label{er_12}}
\end{equation}
Equations (\ref{er_5}), (\ref{er_11}) and (\ref{er_12}) are the central
results of this section.  For $r=0$, these equations reduce to those obtained in~\cite{ScTr04}. 
We now proceed to analyze the various regimes of behavior of the variance
\begin{equation}
Var= \langle x_t^2\rangle - \langle x_t\rangle^2 
\end{equation}
embodied in these results when $r\ne 0$.
A phase diagram associated with this discussion is shown in Fig~\ref{fig:phase}.

\section{Analysis}
\label{analysis}

The most interesting result of this calculation is, of course, the
occurrence of all regimes of behavior, subdiffusive, diffusive, and superdiffusive.  However, there
are additional points to be emphasized because they are not necessarily intuitively obvious.

1. For $\gamma=0$ or $p=q$ (unbiased memory), the mean square displacement
increases sublinearly with time since the exponent $1-r$ is always less than unity for nonzero $r$ values. 
In this case, the mean displacement is independent of time, and so the variance increases
sublinearly. The behavior is therefore subdiffusive except at $r=0$, where it is diffusive. This
is an interesting, perhaps even
counterintuitive result, which says that even the smallest probability of remaining at a given site
at each step, with no other ostensible asymmetry toward the site first stepped on, leads to
subdiffusive behavior.

2. For any fixed $\gamma <0$ we noted earlier that the mean position, which is necessarily one site
away from the origin at the first step, decreases with time. The mean square displacement 
increases sublinearly with time as $t^{1-r}$ when $r>0$.  The variance of a rebellious walker who has even the
smallest probability of staying at a site at each step is thus simply subdiffusive.

3. For fixed $\gamma>0$ the behavior is more varied.  For $0 < \gamma < 1/2$ the 
mean square displacement grows sublinearly with time, as does the square of the mean displacement 
(if it is not zero to begin with).
The resulting variance thus grows subdiffusively for all values of the parameters subject to
this condition, even though the walker more often than not follows its previous randomly chosen step.
However, there are two distinct forms of subdiffusive growth, separated by the line $r=1-2\gamma$  
($p=3q$). These two distinct forms correspond to the dominance of the first or second terms in
Eq.~(\ref{er_11}).  In one, subdiffusion is again caused by the possibility at each step that the
walker may not move. In the other, the walker's rebellion pulls it back.

4. For $\gamma=1/2$ and $r\neq 0$ the growth of the mean square displacement is dominated by the first term in 
Eq.~(\ref{er_11}), $\sim t$. For $\gamma<1/2$ and $r=0$ it is dominated by the second term, and
again grows as $\sim t$.  Together with the contribution of the mean displacement, if present, this
leads to a variance that grows linearly with time, that is, the motion is diffusive. 
The diffusion coefficient is given
by
\begin{equation}
D=\frac{1}{3-4p}-\frac{2(2s-1)^2}{\pi}.
\end{equation}

5. The point $\gamma=1/2$, $r=0$ (where the two red lines meet in color) the motion is marginally
superdiffusive, that is, the variance grows as $Var \sim t\ln t$.

6. When $\gamma>1/2$ the behavior is superdiffusive.  Both the mean displacement (if $s\neq 1/2$)
and the mean square displacement contribute to this behavior of the variance.  

7. The analysis is considerably more interesting if instead of following constant $\gamma$
lines, as we have done
above, we follow the behavior at constant $r$, allowing $p$ and $q$ to vary.  For small $r$ the
figure shows that we cover all regimes of behavior as $p$ increases relative to $q$ ($\gamma$
increased), namely, the two forms of subdiffusive behavior, diffusive behavior, and superdiffusive behavior.
However, if $r$ exceeds the value $1/2$, then we observe only subdiffusive behavior - the tendency
not to move is too strong to allow anything else. 

8. The analysis is also very interesting if instead we follow the behavior at a given value of $p$,
letting $q$ and $r$ vary.  This is shown by the two inclined lines in the figure.  The dashed line
is for $p=0.3$ and the behavior remains subdiffusive throughout.  The persistence is too low to
allow for superdiffusive or even diffusive behavior.  On the other hand if the persistence parameter
is sufficiently large, as in the solid inclined line ($p=0.625$), we sweep all of the behaviors as
$q$ and $r$ are varied subject to this constraint. The value that separates the two regimes is
$p=1/2$.

\section{Conclusion}
\label{conclusion}

We have thus presented a random walk model with memory that exhibits all three forms of
asymptotic behavior as the parameters of the model are varied. The characterization has been
analytical, in terms of the first and second moments of the motion. The model, which is inspired by
one introduced earlier~\cite{ScTr04,CrSiVi07}, has only three simple parameters. One is a
parameter that characterizes the very first step of the walk ($s$).  A second is the probability that the
next step of the walk copies the direction of a randomly chosen earlier step ($p$).  The third is the
probability that the next step performs the opposite motion as that of the randomly chosen earlier
step ($q$).  A fourth, which is actually the crucial new parameter of the model, is the probability
$r$ that the walker simply does not move at the next step.  It is constrained by the others via the
conservation condition $p+q+r=1$. 

The model can be extended in a number of directions.  For instance, one
can explore changes in the nature
of the memory. It is also interesting to calculate quantities other than the first two moments,
perhaps even the full distribution and statistics of extrema.  Perhaps more
interesting would be an adjustment of the model so that it can provide insights into real behaviors
that exhibit these different regimes. 

\section*{ACKNOWLEDGMENTS}
This work was supported in part by the National Science Foundation under grant No. PHY-0855471.


\end{document}